\documentclass[sigplan,10pt,preprint,nonacm,screen]{acmart}
\acmConference[PLDI'21]{ACM SIGPLAN Conference on Programming Languages}{June
20--25 , 2021}{Virtual Conference}
\acmYear{2021}
\acmISBN{}
\acmDOI{}
\startPage{1}
\setcopyright{none}

\usepackage{amsmath,amsfonts}
\usepackage{graphicx}
\usepackage{listings}
\usepackage{float}
\usepackage{subcaption,booktabs}
\usepackage{xcolor}

\usepackage{tabularx,booktabs}
\usepackage{pifont}
\newcommand{\cmark}{\ding{51}}%
\newcommand{\xmark}{\ding{55}}%

\lstdefinelanguage{MLIR}
{
  keywords=[1]{
    add,addf,affine,affine.apply,affine,
    alloc,br,cmpi,constant,delay,dim,dma_start,dma_wait,
    for,func,hir,affine.load,mulf,affine.store,load,store,
    tf,reduce,reshape,matmul,reduce_sum,mhlo,while,Conv2D,xla_hlo,xla_lhlo,return,compare,iota,dynamic_slice,dyname_update_slice,Add,
    mem_read, mem_write, yield, delay, by, call, at, step, offset, to},
  keywordstyle=[1]\color{brown}\ttfamily\bfseries,
  keywords=[2]{
    memref,tensor,vector,i1,i32,f32,f64,i64,tuple,const
  },
  keywordstyle=[2]\color{cyan}\bfseries,
  keywords=[3]{
    func, with, affine_map, affine_set, def, dense, step, tstep, iter_time,
    offset, at
  },
  keywordstyle=[3]\color{magenta}\bfseries,
  morecomment=[l]{//},
  commentstyle=\color{blue},
  columns=flexible,
  mathescape,
  tabsize=2,
  basicstyle=\scriptsize\ttfamily
}

\begin{document}

\title{
HIR: An MLIR-based Intermediate Representation for Hardware Accelerator
Description}


\author{Kingshuk Majumder}
\affiliation{
  \department{Computer Science and Automation}      
  \institution{Indian Institute of Science}            
  \city{Bangalore}
  \country{India}                    
}
\email{kingshukm@iisc.ac.in}          

\author{Uday Bondhugula}
\affiliation{
  \department{Computer Science and Automation}      
  \institution{Indian Institute of Science}            
  \city{Bangalore}
  \country{India}                    
}
\email{udayb@iisc.ac.in}          

\begin{abstract}
The emergence of machine learning, image and audio processing on edge devices
has motivated research towards power efficient custom hardware accelerators.
Though FPGAs are an ideal target for energy efficient custom accelerators, the
difficulty of hardware design and the lack of vendor agnostic,
standardized hardware compilation infrastructure has hindered their adoption.

This paper introduces HIR, an MLIR-based intermediate representation (IR) to
describe hardware accelerator designs. HIR combines high level language
features, such as loops and multidimensional tensors, with programmer defined
explicit scheduling, to provide a high-level IR suitable for DSL compiler
pipelines without compromising control over the micro-architecture of the
accelerator. HIR's explicit schedules allow it to express fine-grained,
synchronization-free parallelism and optimizations such as retiming and
pipelining. Built as a dialect in MLIR, it draws from best IR practices learnt
from communities like those of LLVM. While offering rich optimization
opportunities and a high level abstraction, HIR enables sharing of
optimizations, utilities and passes with software compiler infrastructure.

Our implementation shows that the code generation time of the HIR code generator
is on average 1112$\times$ lower than that of Xilinx Vivado HLS on a range of
kernels without a compromise on the quality of the generated hardware. We
believe that these are significant steps forward in the design of IRs for
hardware synthesis and in equipping domain-specific languages with a productive
and performing compilation path to custom hardware acceleration.

\end{abstract}

\ccsdesc[500]{Hardware~Hardware description languages and compilation}

\keywords{HDL, HLS, MLIR, Verilog, accelerator, FPGA}

\maketitle

\section{Introduction}
\begin{table}[tb]
        \centering
        \begin{tabularx}{\columnwidth}{  l c c c }
            { }                       & \textbf{HDLs}  & \textbf{HLS} & \textbf{HIR}    \\
            \midrule
            Predictable performance          & \cmark        & \xmark & \cmark\\
            \midrule
            Predictable hardware      & \cmark \cmark & \xmark & \cmark\\
            \midrule
            Blackbox modules          & \cmark        & \xmark & \cmark\\
            \midrule
            Sequential execution      & \xmark        & \cmark & \cmark\\
            \midrule
            Deterministic parallelism & \cmark        & \xmark & \cmark\\
            \bottomrule
        \end{tabularx}
        \caption{Comparison of HIR with HLS and HDLs as a target language for
        DSLs.
        \label{tbl:hir_comparison}}
				\vskip -12pt
\end{table}

The growing compute demands of machine learning and other high performance
computing (HPC) domains coupled with the need for power efficiency on edge
computing devices has motivated the increased use of specialized hardware
accelerators.
In the last decade, GPUs have become a de-facto standard for data parallel HPC
workloads. One of the enabling forces behind the mass adoption of GPUs in
mainstream computing is the steady improvements in its programmability.
Both CUDA and OpenCL enabled GPUs for more general purpose programming.  DSLs
such as TensorFlow have helped further in improving the programmability while
simultaneously allowing new kinds of optimizations across multiple kernels. In
short, the continuous improvements in programming abstractions and growing
ecosystem of libraries and DSLs have helped GPUs cater to the ever growing
demands of the HPC community.

Although custom accelerators on reconfigurable computing platforms such as FPGAs
are able to achieve high power efficiency~\cite{zhang16iccad} and performance,
the difficulty of hardware design is seen as a major roadblock for FPGAs. High
level synthesis~\cite{canis2011legup} offers a promising approach towards making
custom accelerator design more approachable. DSLs can offer an even higher level
of abstraction to the algorithm developers and benefit from being able to employ
the right high-level synthesis (HLS) pipelines.

MLIR~\cite{mlir21cgo,mlir2020arxiv}, (multi-level intermediate representation)
is a new compiler infrastructure designed to serve the needs of high-level
domain-specific as well as general purpose programming languages and models at
one end,
and hardware including custom accelerators at the other. Early promising results
have been reported on software code generation as well as on building or
migrating compiler stacks to
MLIR~\cite{uday2020arxiv,mlir-talks,chelini21cgo,jin2020arxiv}.

Numerous domains-specific as well as general-purpose languages and tools
supporting HLS have been built over the past two decades by the academia and
industry~\cite{auerbach12dac,hegarty2014darkroom,rigel,reiche14codes,chugh16pact,vivado,hdl-coder,dase06ieeecs,najjar03computer}.
Nearly all electronic design automation vendors now provide suites supporting
HLS, examples of which include Xilinx Vivado, Cadence C-to-silicon, Synopsys
Synphony, and Mentor Graphics Catapult HLS. A comprehensive survey of various
HLS approaches was conducted by Bacon et al.~\cite{rabbah13acm}.  All of these
HLS efforts devised their own intermediate representations for their purposes,
thus duplicating and re-implementing a lot of the representation and
transformation infrastructure that could otherwise have been shared. The problem
here is thus worse than that faced by the software compiler community including
that faced by domains like deep learning programming systems. The availability
of an open IR standard that is modular and reusable, the same way as
LLVM~\cite{lattner2004llvm}, would bring similar benefits to the HLS compilation
community, and avoid re-engineering core IR infrastructure. Interestingly,
CIRCT~\cite{circt} is an LLVM sub-project and an ongoing umbrella initiative to
use and adapt MLIR and its methodology for HLS: it brings a compiler style
approach to a field where programming and debuggability experience is vastly
different.  Our work here shares the same motivation and end goals, and is
complementary to existing work in CIRCT.

Existing intermediate representations (IRs) for high-level synthesis have not
been built in a rigorous manner the same way as standard compiler IRs such as
LLVM~\cite{lattner2004llvm} and MLIR~\cite{mlir21cgo,mlir2020arxiv}. The latter
draw from best practices learnt from over a decade, and aim to improve compiler
developers' productivity at the expense of significant initial investment in
engineering IR infrastructure. Such engineering and investment purely into IR
infrastructure is made even much before a compiler is built out of it. Our work
in this paper is also along these lines: it focuses on the representation and
the convenience it brings to a hardware developer as opposed to reporting a
specific end-to-end compiler built from the proposed representation.

Our contributions in this paper can be summarized as follows:
\begin{itemize}
  \item{} We design an intermediate representation, HIR, based on the MLIR
    infrastructure, that allows high
    level specification of a hardware design.
  \item{} We introduce support for explicitly specifying the schedule of
    operations in HIR. This makes it easy to exploit fine grained
    parallelism available in the hardware.
  \item{} We build a code generator that translates HIR into
    synthesizable Verilog. The code generator automatically generates the
    controller to realize the schedule specified in HIR.
  \item{} We implement a schedule verification pass in order to automatically
    detect scheduling errors.
  \item{} We implement various optimization passes in the compiler to improve
    the generated hardware design.
\end{itemize}

The rest of this paper is organized as follows. Section~\ref{sec:motivation}
describes our motivation in greater detail. Section~\ref{sec:fpga-overview}
provides an overview of FPGAs and hardware design. Section~\ref{sec:hir}
describes the
HIR representation we propose followed by its advantages in
Section~\ref{sec:advantages}, and its verification and optimization
infrastructure in
Section~\ref{sec:auto-opt} and Section~\ref{sec:manual-opt}.  Related work is
discussed in Section~\ref{sec:related-work}, and an evaluation is presented in
Section~\ref{sec:evaluation}.


\section{Motivation}
\label{sec:motivation}

Traditionally, domain-specific languages have targeted either high level
synthesis (HLS) or hardware description languages (HDLs) as the output. With
HLS, the DSL designer is able to generate code in a high level language such as
C/C++. This makes it easy to implement the code generator. HLS compilers also
provide various mechanisms (such as pragmas) to tune the generated hardware for
different performance and area tradeoffs.  In contrast, HDLs are low level but
offer complete control over the generated hardware. The DSL designer is able to
manually optimize the design's critical sections and timing errors are easier to
debug than with HLS. Another advantage of HDLs is that they are standardized and
most tools understand Verilog or VHDL.

An IR for hardware description has the following desirable properties:
\begin{itemize}
  \item{Quality of Result (QoR):} A description of a hardware design should have
    a guaranteed minimum QoR (latency, throughput and area). Verilog, being a
    low level language provides this guarantee but HLS languages heavily depend
    on the compiler to achieve a certain performance. Compared to HDLs, HLS
    designs have lower performance portability across different compilers.
  \item{Ease of optimization:} It should be easy for the compiler to find
    optimization opportunities in the design. HIR designs define the schedule
    explicitly, which helps in optimization. For example, if an distributed RAM
    is defined as simple dual port but the read and write operation's schedules
    do not overlap, we can replace it with a single port RAM to save resources.
    HDLs make it difficult to extract the schedule from a given design which
    makes these kind of optimizations much harder. Traditionally, HLS compilers
    perform these kind of higher level optimizations since the compiler itself
    schedules the HLS design.
  \item{Interfacing with external modules:} The IR should be able to interface
    with external IPs without much overhead. This is essential to leverage
    existing libraries. HLS languages require various handshake protocols to be
    implemented for using outside blackbox IPs in the HLS design.
  \item{Predictable design:} The output hardware should be predictable from the
    description written in the IR. As HLS compilers perform different kinds of
    optimizations during the binding and scheduling phase, it is harder to
    predict the resource usage of an HLS design.
  \item{Smaller abstraction gap:} Many DSLs target HLS languages because it is
    much more difficult to describe the design in HDLs. An IR that can be used
    by DSLs needs to be high level. It is desirable to have basic programming
    constructs such as loops. Most software algorithms contain a mix of
    sequential and parallel regions. Thus, while translating the algorithm to a
    hardware implementation, the IR should provide language constructs to easily
    express both sequential and parallel algorithms. The abstraction of HDLs is
    very close to hardware and hence, like hardware, all operations in HDLs are
    parallel.  Running dependent operations sequentially requires additional
    effort from the developers' side such as creation of state machines. HLS
    languages often can only express non-deterministic parallelism. As a result,
    parallel operations may need extra synchronization logic.
\end{itemize}

\section{Overview of FPGAs}
\label{sec:fpga-overview}
In this section, we discuss about accelerator design and the different kinds of
hardware resources available in FPGAs. We will use the nomenclature of Xilinx
FPGAs, but the concepts generalize to other FPGAs as well.

FPGAs are a fundamentally different kind of hardware from CPUs and GPUs. An FPGA
is a grid of simple building blocks such as small lookup tables, SRAMs,
registers and ALUs (DSP slices). FPGAs do not execute instructions and thus, do
not have an ISA. Instead, FPGAs can implement arbitrary digital circuits by
connecting the hardware blocks using a programmable on-chip routing matrix.

FPGAs offer multiple types of on-chip memory. For example, Xilinx FPGAs contain
registers, distributed RAMs and block RAMs. These memories vary in their size
and power consumption. Registers hold just one bit of data. Smaller buffers are
implemented using distributed RAMs, and block RAMs offer larger capacity at the
expense of greater power consumption. Each RAM has a fixed number of read/write
ports. These ports can be used independently for parallel memory accesses.
FPGAs lack any kind of caches. Accelerators implement custom memory hierarchy to
cache the frequently accessed data in the on-chip memory buffers for fast,
energy efficient access.

FPGA vendors usually provide custom synthesis tools to target their FPGAs. These
tools can take a hardware design as input and perform optimizations, placement
and routing to output a configuration file that can be used to program the FPGA.
Any external compiler must communicate with the synthesis tools to target the
FPGA. This often means that the external compiler must output the design in a
hardware description language (HDL) that is supported by the synthesis tool. The
two most widely supported HDLs are Verilog and VHDL.

A hardware design is usually synchronized using a clock, i.e., all state changes
happen only at discrete clock ticks. Unlike CPUs and GPUs, FPGAs provide
deterministic fine-grained parallelism. All operations in the hardware design
execute every cycle unless explicitly constrained. Hardware designs often use
finite state machines (FSMs) to choose which operations are enabled at each
clock cycle. Thus, in contrast to software development, a design must use extra
logic to ensure sequential execution while parallel execution is the default.

An accelerator design in a hardware description language, such as Verilog or
VHDL, is usually broken into data paths and FSMs. The data paths implement the
computation and the FSMs control when each operation in the data path executes.

\section{HIR}
\label{sec:hir}

The HIR intermediate representation is implemented as a dialect in the
MLIR~\cite{mlir21cgo} compiler infrastructure. As such, it inherits all the
usual benefits provided by the core MLIR infrastructure. This includes having a
round-trippable and human readable textual representation that could be parsed,
printed, and verified~\cite{mlir2020arxiv}.  All our HIR operations have a
custom pretty-printed form for readability and for the convenience of compiler
developers.
Listing~\ref{code:hir_transpose} shows a sample design in HIR dialect to
transpose a matrix. HIR borrows its syntax from software programming languages.
Like LLVM, all variables in HIR are SSA variables.

The hardware implementation of any algorithm can be broken down
into three orthogonal components :
\begin{itemize}
  \item{} The high level \textit{algorithm}: this can be captured by any software
    programming language.
  \item{} The \textit{schedule} of each constituent operation, i.e., the clock cycle
    at each an operation is executed.
  \item{} \textit{Binding} operations and states (variable/arrays) to available
    hardware resources such as multipliers, lookup tables, registers and
    different kinds of on-chip SRAMs.
\end{itemize}

\begin{lstlisting}[language=MLIR,float=b,frame=bt,breaklines=true,captionpos=b,caption=HIR
code for matrix transpose.,label={code:hir_transpose}]
hir.func @transpose at %t(
  %Ai :!hir.memref<16*16*i32, r>,
  %Co : !hir.memref<16*16*i32, w>) {

  %0 = hir.constant 0
  %1 = hir.constant 1
  %16 = hir.constant 16

  hir.for %i : i32 = %0 : !hir.const to %16 : !hir.const
    step %1:!hir.const iter_time(%ti = %t offset %1 ){

      %tf = hir.for %j : i32 = %0 : !hir.const to %16 : !hir.const
        step %1:!hir.const iter_time(%tj = %ti offset %1){
          %v =  hir.mem_read %Ai[%i, %j] at %tj
              : !hir.memref<16*16*i32, r>[i32, i32] -> i32
          %j1 = hir.delay %j by %1 at %tj: i32 -> i32
          hir.mem_write %v to %Co[%j1, %i] at %tj offset %1
              : (i32,!hir.memref<16*16*i32, w>[i32, i32])
          hir.yield at %tj offset %1
      }

      hir.yield at %tf offset %1
  }
  hir.return
}
\end{lstlisting}

Hardware description languages require the developer to bind operations and
manually schedule them using state machines. High level synthesis takes the
other extreme approach where the developer only specifies the algorithm and the
compiler performs binding and scheduling. The developer can influence the
scheduling and binding via annotations but does not have direct control over it.
For example, in Vivado HLS~\cite{vivado}, the programmer can add loop pipelining
annotations
with specific initiation interval but compiler can reject it if it can't
statically prove that pipelining will not violate loop carried dependencies.

HIR provides complete control over the schedule of the design while still
providing high level abstraction. HIR's primary insight is that the complexity
of hardware description in HDLs partly come from manually implementing the
schedule using state machines. In HIR dialect, the programmer only
\textit{describes} the schedule and the compiler automatically generates the
control logic. Thus the abstraction of HIR is in between HDLs and HLS (where
there is no scheduling information in the design and the compiler tries to
determine an optimal schedule).

\begin{table}
  \centering
  \begin{tabularx}{\columnwidth}{  l p{0.6\linewidth}}
    \toprule
    \textbf{Category} & \textbf{Examples} \\
    \midrule
    Data types        & {\it i32, i1, f32, hir.memref}\\
    \midrule
    Control flow ops  & {\it hir.func, hir.for, hir.unroll\_for, hir.return,
    hir.yield}\\
    \midrule
    Compute ops       & {\it hir.add, hir.mult} \\
    \midrule
    Memory access ops & {\it hir.mem\_read, hir.mem\_write}   \\
    \bottomrule
  \end{tabularx}
  \caption{Data types and different types of operations in HIR.
  \label{tbl:hir_overview}}
  \vskip -10pt
\end{table}

\subsection{Operations}
\label{sec:hir_operations}
An `operation' is the building block of all MLIR programs. Each operation can
have multiple inputs, outputs, regions and constant attributes. A region defines
a new lexical scope enclosed in curly braces. Control flow operations such as
functions, conditionals and loops require region arguments. Attributes associate
constant metadata with an operation. The HIR dialect defines three types of
operations.

\textbf{Control flow operations} help in expressing accelerator designs at a
high level of abstraction.
Most HDLs and hardware IRs support conditional statements and generate loops
(equivalent to an \textit{unroll-for} loop in HIR) in their synthesizable
subset. In addition to these, HIR also adds \textit{for} loop in the language.
The compiler automatically generates state-machines to implement the control
flow in hardware.
Listing~\ref{code:hir_transpose} shows the syntax of the \textit{for} loop. The
loop takes a lower bound, an upper bound, a step, and start time as its inputs.
It also defines two SSA variables that are accessible inside the loop body: the
loop induction variable and a time variable to represent the start of an
iteration. These two variables have different values for each iteration of the
loop. Operations inside the loop do not have access to any time variable other
than the iteration start time. This ensures that each operation inside the body
can only be scheduled after the loop guard is checked.

\textbf{Compute operations} calculate output values given the inputs. This
includes arithmetic and logical operations, bit slicing ops and function calls.
The `call' operation can be used to execute another HIR function or to
communicate with an external Verilog module.

\textbf{Memory access operations} such as \textit{hir.mem\_read} are used
to access memref types. Memory access operations are constant delay operations.
All memory writes take one cycle. Memory reads may take zero or one cycle
depending on whether the memref is implemented using registers or on-chip
buffers.

\subsection{Time variable and schedules}
\label{sec:hir_schedule}
A key contribution of the HIR dialect is the concept of time variables and
schedules.  A time variable represents a specific time instant within its
lexical scope. This time instant is either related to some event, such as a
function call, or represents a constant delay from another time variable.

In HIR, the keyword \textit{`at'} is used to express that an operation starts at
the time instant represented by a given time variable.  For example, in
Listing~\ref{code:hir_transpose}, the time variable $\%t$ represents the time at
which the \textit{transpose} function is called. The schedule for all operations
within the function's body are described with respect to this time variable.  The caller
would schedule the function call with respect to its own start time. This recursively
defines the concrete schedule of all operations in all functions in the call
graph starting from the top-level function.

The start of each iteration of a \textit{for} loop is also associated with a
time variable. The $i$-loop in Listing~\ref{code:hir_transpose} defines the time
variable $\%ti$ to represent the start of an iteration. This time variable is
local to the loop body, and for each loop iteration, the time variable represents
a different time instant. At the first iteration, $\%ti = \%t+1$, i.e., the loop
starts its execution at time $\%t+1$ (represented as `$\%t$ offset $\%1$'). The
\texttt{hir.yield} operation decides the time at which the next iteration
starts. The $i$-loop is sequential and the next iteration starts after the
inner $j$-loop completes.

The {\tt yield} operation is used for loop pipelining. The j-loop body in
Listing~\ref{code:hir_transpose} yields one cycle after an iteration starts, i.e.,
if the first iteration is at $\%tj=\%ti+1$, the second iteration starts at
$\%tj=\%ti+2$ and so on. The {\tt yield} operation does not end the current iteration.
Each iteration of the j-loop takes two cycles to complete (the mem\_write
operation starts at $\%tj+1$ and takes one cycle to complete). Thus, the
execution of different iterations of the j-loop overlap, i.e., the j-loop is
pipelined. Although multiple iterations can execute at the same time, the loop
iteration time variable will still hold the correct iteration.

As the schedule of each operation is explicitly specified using time variables,
the textual order of operations does not have any effect on the functionality.
For example, if the {\tt yield} operation in the j-loop of the matrix transpose example is
moved to the beginning of the loop body, it will still not change the behavior
of the circuit. The only constraint on the textual order of the operations is
that the definition of an SSA variable must dominate all its uses.

\subsection{Primitive types and HIR constants}
HIR supports primitive data types such as arbitrary bit-width integers and
floats. Each SSA variable of primitive type is defined only at a specific time
instant within its lexical scope. For example, the $\%v$ variable in
Listing~\ref{code:hir_transpose} contains a valid value only at time $\%tj+1$
(since mem\_read takes one cycle). Constant type (\textit{hir.const} in the
dialect) represents a constant integer in HIR.

\subsection{Memref data type in HIR}
All the available memory resources in hardware are represented via the
\textit{hir.memref} data type. A memref can be viewed as a pointer or reference
to a multidimensional tensor. The tensor may be placed in an array of buffers
(such as distributed or block RAM) or registers. The memref type abstracts its
implementation details away and provides a uniform interface for memory access.

A memref defines the dimensions of the tensor, the data type of its elements, and
its access permission. Memref types can have either a
read-only, write-only or a read-write access permission. A single tensor may be associated
with multiple memrefs pointing to it. Each memref pointing to the tensor
represents a memory port. The maximum number of memrefs that can simultaneously
point to a tensor is fixed by the available number of ports. For example,
block RAMs in Xilinx FPGAs are dual ported. Thus, a tensor that is held in
the block RAM can have two memrefs pointing to it.

Each dimension of a memref is either a packed dimension or a distributed dimension.  The
tensor elements that have the same indices for the distributed dimensions go to
the same buffer. Elements whose address differs in a distributed dimension go to
different buffers.  Distributed dimensions can only be indexed using compile
time constants (i.e.,  \textit{hir.const} type). Multiple parallel read/write
operations on the same memref is valid only if either the address is same for
all the conflicting operations or each access is to a different buffer, i.e., the
addresses differ in at least one distributed dimension. For unrestricted parallel
access, a design may use multiple memrefs pointing to the same tensor.

\subsection{Undefined behavior}
HIR also borrows the concept of undefined behavior from software programming
languages. The HIR compiler makes the following assumptions:
\begin{itemize}
  \item{} Memory accesses remain within bounds.
  \item{} Lower bound of a for-loop is never greater than the upper bound.
  \item{} There will never be multiple accesses to a memref in the same clock
    cycle unless they occur at the same address or at least one distributed
    dimension index is different. This is because each memref corresponds
    to one port of an on-chip buffer.
  \item{} A new instance of a for-loop is not scheduled unless the previous
    instance has completed all iterations. For this reason, the inner loop in
    listing~\ref{code:hir_transpose} is pipelined but the outer loop is
    sequential.
  \item{} All {\tt mem\_read} operations happen on initialized memory, i.e., the memory
    must be written-to before reading a value from it. Each call to a function
    resets all memory elements (such as registers and RAMs) instantiated in the
    function to uninitialized state. The HIR language does not support
    persistent state (equivalent to static scope in C) across function calls.
\end{itemize}

Violation of any of the above assumptions is treated as undefined behavior.
This assists the HIR compiler in two ways.  First, it allows the code generator
to automatically insert assertion statements in the generated Verilog to guard
against violation of the above assumptions.  These assertions can then be used
to verify that the circuit is not malfunctioning via simulation or formal
verification techniques. Traditional HDLs describe hardware at a lower level of
abstraction and thus can not describe these types of undefined behaviors.  For
example, in Verilog, the programmer would manually implement an if-else
statement (which synthesizes to a multiplexer), to select between multiple
addresses and read enable signals, when there are multiple readers.  If these
readers attempt to read at different addresses from the same memory port in the
same clock cycle, then it is very likely that the design is malfunctioning. But
to the Verilog compiler, it is perfectly valid for multiple branches of the
if-else statement to be true -- only the first branch is selected. There is not
enough information for the HDL compiler to understand the programmer's intent
in such situations. Thus, often, the programmer would have to manually insert
these assertions.  Due to HIR's higher level abstraction, it is possible to
automatically add assertions to check that only one of the memory operations is
to be executed at a given cycle.

The second usual advantage of undefined behaviors is that the optimizer can
exploit this information in order to improve the circuit. For instance, if a
variable is used to index into an array, the compiler can assume that the
variable will never contain a value greater than the size of the array. This
enables precision reducing optimizations.

\subsection{Lowering to hardware}
\begin{table}[tb]
  \centering
  \begin{tabularx}{\columnwidth}{  l X }
    \toprule
    \textbf{HIR } & \textbf{Verilog}    \\
    \midrule
    Functions                        & Verilog modules \\
    \midrule
    Primitive types                  & Wires     \\
    \midrule
    memrefs                          & Block RAMs, distributed RAMs
    \newline and registers. \\
    \midrule
    Integer arithmetic              & Non-pipelined arithmetic in
    \newline Verilog   \\
    \midrule
    Delay                            & Shift registers   \\
    \midrule
    For loops                        & State machines \\
    \midrule
    Schedules                        & State machines \\
    \bottomrule
  \end{tabularx}
  \vskip 5pt
  \caption{Hardware mapping for each HIR language construct.
  \label{tbl:hardware_mapping}}
  \vskip -10pt
\end{table}

HIR is a thin abstraction on top of hardware. This is intentional in order to
support low overhead and predictable hardware generation. HIR's code generator
outputs Verilog which is understood by nearly every hardware toolchain.
Table~\ref{tbl:hardware_mapping} shows the hardware that is instantiated by our
code generator for each HIR language construct.

Functions in HIR map to modules in Verilog. Basic integer arithmetic operations
such as integer addition, subtraction and multiplication map to the
corresponding operations in Verilog. Note that all these operations are
combinatorial, i.e., they are not pipelined. Delay operations are mapped to
shift registers.

Variables of primitive types such as {\it i32} and {\it f32} map to wires in
Verilog. The \textit{hir.memref} type is implemented as a combination of
address, `read enable', `write enable' and data buses depending on whether it is
a read-only, write-only or a read-write port. The multi-dimensional array that a
memref refers to is implemented using either registers or on-chip memory buffers
such as block RAMs.  The order of packed dimensions decides the layout of data
within each RAM.  The distributed dimensions produce a banked RAM design.  The
code generator adds circuitry to calculate the linear address and the signals to
select the RAM within the bank based on the input addresses.

The schedule is implemented in hardware as state machines that generate control
signals for other operations. The state machine also generates clock enable
signals for all pipeline registers. All Verilog modules generated from HIR
functions contain a start signal that the caller's state machine uses to signal
a function call.

\section{Advantages of HIR}
\label{sec:advantages}
In this section, we discuss the advantages of using HIR as an intermediate
language for DSL compilers.

\subsection{Explicit scheduling}
\label{sec:explicit_scheduling_advantage}

We can broadly group existing languages (and their IRs) for hardware description
into two categories: languages that precisely describe the schedule of
operations in the hardware design, and languages that rely on the compiler to
generate a schedule. HDLs such as Verilog or VHDL,
Chisel~\cite{bachrach2012chisel} and IRs such as LLHD~\cite{schuiki2020llhd},
FIRRTL~\cite{izraelevitz2017firrtl} fall under the first category, while
high-level synthesis tools such as Vivado HLS~\cite{vivado},
Bluespec-Verilog~\cite{nikhil2004bluespec} belong to the second category.

Languages that precisely capture the schedule usually require the programmer to
manually implement state machines. These state machines then generate the
control signals that determine the order of execution of operations. Without
these control signals, every operation in the hardware will execute every cycle.
The problem with this approach is that the scheduling information is hidden
inside the controller, and an optimization requiring the schedule needs to
rediscover this information from the controller description. Also, any change in
the schedule requires reimplementing the controller. In HIR, the programmer
specifies the relative time of execution of each operation and the compiler
automatically generates the required controllers. This allows HIR to do
optimizations that require the scheduling information. For example, loop
pipelining requires control over the schedule of execution of each iteration of
the loop. Similarly, retiming and operator chaining are a few other
optimizations that modify the schedule.

An explicitly defined schedule makes it easy to modify the schedule to optimize
the design. A similar modification in HDL would require updating the controllers
as well. An added advantage of this approach is that the compiler can
automatically detect many scheduling errors (see
section~\ref{sec:schedule_verification}).

\subsection{High level design}
HIR borrows control flow constructs such as loops, function calls and
conditional statements directly from imperative programming languages. The
explicit schedules make it easy to enforce sequential execution when there is a
dependence between operations. These features make it easy to convert software
algorithms into hardware designs.  An HIR design can be optimized using HLS
optimization techniques such as loop pipelining and overlapped execution. In
addition, HIR offers more control over the generated hardware when compared to
HLS. For instance, an HIR design completely controls the pipelining and
scheduling of individual operations. This additional control can be used for
optimizations such as adding pipeline registers and retiming to improve the
frequency of operation.

\subsection{Deterministic parallelism}
Unlike HDLs, HLS languages cannot express deterministic parallelism. This is
primarily because they do not explicitly schedule operations, and the languages
used by HLS tools are usually software programming languages with constructs for
non-deterministic coarse-grained parallelism. This kind of parallelism often
requires some kind of a synchronization mechanism.  For example, in Vivado HLS a
producer and a consumer task can execute in parallel if the producer is
transferring its outputs to the consumer via streams (implemented as FIFOs in
hardware). The consumer task waits for the next data to be available on the
stream in order to continue its execution. This type of synchronization adds
overhead in the hardware design. The producer task need to assume that the FIFO
may be full and thus the generated circuit needs a way to pause (referred to as
back-pressure). Similarly, the consumer needs to provision for the situation
when the FIFO is empty. If the two tasks are working in lock-step, i.e., every
fixed number of cycles with the producer task generating one output and the
consumer task consuming it, then there is no need for synchronization between
the two tasks. Both HIR and HDL languages can express this kind of
deterministic, synchronization-free, task level parallelism.
Section~\ref{sec:manual-opt} shows an example of task level parallelism in a 1-d
stencil expressed in HIR.

\subsection{External hardware modules}
The ability to use externally defined hardware circuits is essential in order to
specialize a design for the given hardware platform. This is because FPGA
vendors provide their own custom libraries for many common circuits. For
instance, Xilinx offers optimized libraries for floating-point arithmetic,
multi-ported RAMs, etc. for their FPGAs. Often these library implementations do
not have their source code available. Thus, porting these libraries to the IR
is not an option.  Additionally, there are several third party libraries that
may need to be reused or inter-operated with.

HIR's ability to capture scheduling information in the function signature makes
it easier to integrate external Verilog modules with HIR's design. In languages
where the schedule is determined by the compiler, external modules usually
require handshake signals. Since HIR allows the programmer to precisely specify
when data is read from or written to a wire or memory, handshake signals are not
required unless the external hardware module does not have a fixed latency.

\subsection{Timing closure}
Often, hardware designs have a target frequency requirement. Timing closure is
the process of iterating over the design to meet the required frequency of
operation. The peak frequency of operation for a design is limited by the delay
in the critical path (longest timing path) of the circuit. This delay is a sum
of the wire delays (time for signals to propagate through wire) and delay
introduced by the logic in the critical path. Improving the frequency of the
design involves breaking down the critical path by introducing pipeline stages.
All HDLs and IRs that control the scheduling of operations can introduce
additional pipeline stages to meet timing requirements. HIR allows pipelining
using delay operations and modifying the schedule. HIR has an additional
advantage that the compiler can detect pipeline imbalance errors (discussed in
section~\ref{sec:schedule_verification}).

We also use MLIR's ability to track location information in order to generate a
mapping between the operations in HIR and the corresponding Verilog code. The
code generation pass prints the location information of the HIR operation
as comments in the generated Verilog output. This helps us in identifying the
critical paths in an HIR design in case of timing failures in the generated
Verilog design.

\subsection{Predictable QoR}
A predictable quality of result (QoR) is essential for a hardware intermediate
representation. This allows DSLs to generate hardware with a predictable
performance and resource usage. Control over resource usage is also important
when the generated accelerator has to share the FPGA with other hardware
designs. For example, the generated accelerator may be deployed as a
part of a larger design containing PCIe controllers, DRAM controllers and soft
CPU cores (CPU implemented on an FPGA).

The performance of a hardware accelerator is determined by the amount of
parallelism exploited by the accelerator, the memory bandwidth and the frequency
of operation. Among these, the latter two are dependent on the target FPGA.
An HIR design contains a description of the schedule of all operations.
This schedule precisely specifies the amount of parallelism in the design.
In HIR, all resources are explicitly instantiated in the design. No extra
hardware resources are required to meet the performance specified in the design.
In contrast, HLS compilers can add extra hardware such as ping-pong buffers and
extra memory ports in order to meet the desired performance.

\section{Automatic Verification and Optimization}
\label{sec:auto-opt}
In this section, we discuss various passes that we implemented in the HIR
compiler for optimizing IR and to detect anomalies in a design.

\subsection{Verifying the schedule of computation.}
\label{sec:schedule_verification}
Additional verification of IR helps language front-ends catch errors
early in the development cycle. It also helps library implementors and DSL
designers write hand-optimized kernels directly in HIR.

HDLs do not have any inherent notion of \textit{logically valid} data in the
language. For example, a wire in Verilog always contains some data. The compiler
cannot determine when this data is valid.  The SSA variables of
primitive types ({\it int} and {\it float}) in HIR specify the time instant
(relative to a time variable) at which they will have a valid value. The
schedule verification pass exploits this \textit{validity} information, and the
explicitly specified schedule of operations, to detect errors at compile time
that cannot be automatically detected in an design specified in an HDL.

\begin{figure}
  \begin{subfigure}{\linewidth}
\begin{lstlisting}[language=MLIR,frame=bt,breaklines=true,captionpos=b
    ]
hir.func @Array_Add at %t (%A:!hir.memref<128*i32, r>,
  %B : !hir.memref<128*i32, r>,
  %C : !hir.memref<128*i32, w>){

  %0 = hir.constant 0
  %1 = hir.constant 1
  %128 = hir.constant 128

  hir.for %i :i8 = %0 :!hir.const to %128 :!hir.const
    step %1 :!hir.const iter_time(%ti = %t offset %1){
        hir.yield at %ti offset %1

        %a = hir.mem_read %A[%i] at %ti
            : !hir.memref<128*i32,r>[i8] -> i32
        %b = hir.mem_read %B[%i] at %ti
            : !hir.memref<128*i32, r>[i8] -> i32

        %c = hir.add (%a, %b) : (i32, i32) -> (i32)

        hir.mem_write %c to %C[%i] at %ti offset %1
            : (i32, !hir.memref<128*i32, w>[i8])
  }
  hir.return
}
\end{lstlisting}
    \caption{HIR design to add two arrays.}
    \label{code:hir_add}
  \end{subfigure}
  \par\bigskip
  \begin{subfigure}{\linewidth}
    \includegraphics[width=\linewidth]{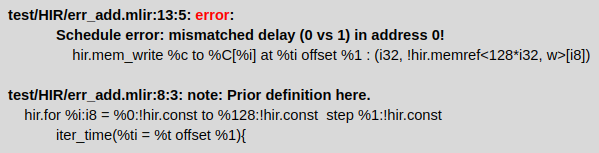}
    \caption{Diagnostic error reported by HIR's schedule verifier.}
    \label{fig:schedule_error}
  \end{subfigure}
  \caption{Example of an HIR design containing a scheduling error.}
  \label{fig:hir_add}
\end{figure}

Figure~\ref{fig:hir_add} shows an example of a wrong schedule. The design
in Figure~\ref{code:hir_add} adds two arrays and writes the output into a third
array. Figure~\ref{fig:schedule_error} shows the error reported by our schedule
verification pass.
The \textit{for} loop has an initiation interval of $1$ as
specified by the \texttt{yield} operation inside the body of the loop. The
\textit{mem\_write} operation takes $\%i$ as an operand at time $\%ti+1$ but the
state machine of the \textit{for} loop generates $\%i$ at time $\%ti$. Since the
\textit{for} loop has an initiation interval of one cycle, when the
\textit{mem\_write} occurs at $\%ti+1$, the $\%i$ variable would have
incremented already.

\begin{figure}
\begin{subfigure}{\linewidth}
\begin{lstlisting}[language=MLIR,frame=bt,breaklines=true,captionpos=b
    ]
hir.func @mac at %t (%a :i32, %b :i32, %c :i32) -> (i32 delay 3){
  %1 = hir.constant 1
  %2 = hir.constant 2

  // Pipelined multiplier.
  %m = hir.call @mult_3stage (%a,%b) at %t
        : (i32, i32) -> (i32 delay 3)

  %c2= hir.delay %c by %2 at %t : i32 -> i32

  %res = hir.add (%m,%c2) : (i32, i32) -> (i32)
  %res1 = hir.delay %res by %1 at %t offset %2 : i32 -> i32

  hir.return (%res1) : (i32)
}
\end{lstlisting}
  \caption{HIR design for multiply-accumulate operation.}
  \label{code:hir_mac}
\end{subfigure}
\par\bigskip
\begin{subfigure}{\linewidth}
  \includegraphics[width=\linewidth]{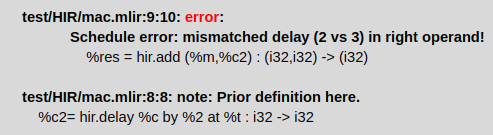}
  \caption{Diagnostic error reported by HIR's schedule verifier.}
  \label{fig:schedule_error_pipeline}
\end{subfigure}
\caption{Example of a pipeline imbalance.}
\label{fig:hir_mac}
\end{figure}

In hardware design, it is common to add extra pipeline stages to a circuit in
order to achieve the desired target frequency. But this may introduce new
errors in the final design. Figure~\ref{fig:hir_mac} illustrates this with an
example design of a multiply-accumulate operation in which a two stage integer
multiplier is replaced with a multiplier that has three pipeline stages. For
the adder to work properly, both its inputs must arrive at the same clock
cycle. The added pipeline stage in the multiplier delays $\%m$ by one cycle
leading to a malfunctioning design.  Since, function signatures in HIR embed
the delays of each input and output value, we can analyze the source code to
find these pipeline mismatches at compile time as shown in
Figure~\ref{fig:schedule_error_pipeline}.

\subsection{Constant propagation, CSE, and strength reduction}
The HIR compiler performs the standard optimizations well-known in the software
compiler domain like constant propagation and common sub-expression
elimination.  These optimizations enable subsequent optimizations and help
reduce resource usage. The optimizer replaces multiplication between loop
induction variables and constants with increments.  This reduces the hardware
usage since integer type multiplication consumes more hardware resources than
addition.

\subsection{Precision optimization}
\begin{table}[tb]
        \centering
        \begin{tabularx}{0.7\columnwidth}{  l c  c }
            \toprule
            { }                     & LUT & FF   \\
            \midrule
            Vivado HLS              & 41  & 92   \\
            \midrule
            Vivado HLS (manual opt) & 7  & 51   \\
            \midrule
            HIR (no opt)            & 32  & 72 \\
            \midrule
            HIR (auto opt)          & 8  & 18 \\
            \bottomrule
        \end{tabularx}
        \vskip 5pt
        \caption{Resource usage of a Matrix Transpose.
        \label{tbl:transpose_resource}}
				\vskip -10pt
\end{table}

Compared to software, where the minimum precision of the compute unit is
determined by the underlying architecture, hardware designs can greatly benefit
from reducing the precision of arithmetic to arbitrarily low levels. The HIR
compiler automatically analyzes the source code to reduce the precision of the
arithmetic operations in certain simple cases and thereby improves resource
usage. Here, we benefit from the higher level description of HIR.  For example,
constant loop bounds help in determining the minimum precision required to
calculate the loop induction variable. To do the equivalent optimization, the
HDL optimizer would have to analyze a state machine.
Table~\ref{tbl:transpose_resource} compares the resource utilization of the
matrix transpose. We manually optimized the HLS design because Vivado HLS was
not able to perform the optimization automatically. The results show that
precision optimization significantly improves resource utilization both in
Vivado HLS (with manual precision reduction) and HIR (with automatic precision
reduction).

\subsection{Delay elimination}
An HIR design may contain several \texttt{hir.delay} operations in order to
maintain the validity of the schedule. Each delay operation is implemented
using shift registers. In order to reduce the use of registers, the compiler
reuses shift registers across {\tt hir.delay} operations. The compiler first
applies a de-duplication pass on the design to remove unnecessary time
variables. Then delay operations using the same time variable and with same
inputs can share the shift registers.

\section{Manual optimizations}
\label{sec:manual-opt} DSL compilers are expected to exploit domain knowledge
to find potential optimization opportunities~\cite{hegarty2014darkroom}. The
intermediate language should allow the DSL compiler to express these
optimizations in the accelerator design so that the compiler backend can
generate the desired circuit. In this section, we discuss different ways in
which an HIR design can be optimized for improved frequency and
parallelization. These optimizations may be manually employed for
hand-optimized kernels or could be used by auto-parallelization frameworks to
emit efficient hardware designs. Automatic parallelization of HIR designs is
itself beyond the scope of this work. Although an intermediate representation is
not meant to be a programming model or a language, the high-level nature of
abstractions and operations provided by HIR make manual optimization by a
designer a useful feature.

\subsection{Loop pipelining}
\begin{lstlisting}[language=MLIR,float=hb,frame=bt,breaklines=true,captionpos=b,
caption={One-dimensional stencil with a pipelined
loop.},label={code:hir_stencil}]
hir.func @stencil_1d at %t(
  %Ai :!hir.memref<64*i32, r>,
  %Bw : !hir.memref<64*i32, w>
) {
  %0  = hir.constant 0
  %1  = hir.constant 1
  %2  = hir.constant 2
  %3  = hir.constant 3
  %64 = hir.constant 64
  %W1r, %W1w = hir.alloc() : !hir.memref<2*i32, packing=[], r>,
                          !hir.memref<2*i32, packing=[], w>
  %valA = hir.mem_read %Ai[%0] at %t
      : !hir.memref<64*i32, r>[!hir.const] -> i32
  %valA1 = hir.delay %valA by %1 at %t offset %1: i32 -> i32
  %valB = hir.mem_read %Ai[%1] at %t offset %1
      : !hir.memref<64*i32, r>[!hir.const] -> i32
  hir.mem_write %valA1 to %W1w[%0] at %t offset %2
      : (i32,!hir.memref<2*i32, packing=[],w>[!hir.const])
  hir.mem_write %valB to %W1w[%1] at %t offset %2
      : (i32,!hir.memref<2*i32, packing=[],w>[!hir.const])
  // Pipelined for loop.
  hir.for %i : i32 = %1 : !hir.const to %64 : !hir.const
    step %1:!hir.const iter_time(%ti = %t offset %3) {
      // Yield instruction starts next iteration after one cycle.
      hir.yield at %ti offset %1
      %v0 = hir.mem_read %W1r[%0] at %ti offset %1
          : !hir.memref<2*i32,packing=[], r>[!hir.const] -> i32
      %v1 = hir.mem_read %W1r[%1] at %ti offset %1
          : !hir.memref<2*i32,packing=[], r>[!hir.const] -> i32
      %iPlus1 = hir.add (%i,%1) : (i32,!hir.const) -> (i32)
      %v =  hir.mem_read %Ai[%iPlus1] at %ti
          : !hir.memref<64*i32, r>[i32] -> i32
      hir.mem_write %v1 to %W1w[%0] at %ti offset %1
          : (i32,!hir.memref<2*i32,packing=[], w>[!hir.const])
      hir.mem_write %v to %W1w[%1] at %ti offset %1
          : (i32,!hir.memref<2*i32,packing=[], w>[!hir.const])
      %r  = hir.call @stencil_opA(%v0, %v1) at %ti offset %1
          : (i32, i32) -> (i32 delay 1)
      %i2 = hir.delay %i by %2 at %ti: i32 -> i32
      hir.mem_write %r to %Bw[%i2] at %ti offset %2
          : (i32,!hir.memref<64*i32, w>[i32])
  }
  hir.return
}
\end{lstlisting}

Loop pipelining is a key optimization in high level synthesis.  In loop
pipelining, the next iteration of the for loop starts before the previous
iteration completes. This allows multiple loop iterations to execute in
parallel. Loop pipelining does not add extra hardware overhead.
Listing~\ref{code:hir_stencil} shows an example of a pipelined loop. Each
iteration of the i-loop in the function "stencilA", takes 3 cycles to complete.
The time to schedule the next iteration of the loop is determined by the
\texttt{hir.yield} operation. In this case the i-loop starts every cycle. The
{\tt yield} operation allows an HIR design to express a constant or a variable
initiation interval.

\subsection{Task level parallelism}
\begin{lstlisting}[language=MLIR,float=bt,frame=bt,breaklines=true,captionpos=b,
caption={Overlapped execution of tasks.},label={code:hir_task_parallel}]
hir.func @task_parallel at %t(
  %Ai :!hir.memref<64*i32, r>,
  %Co : !hir.memref<64*i32, w>) {
  %Br, %Bw = hir.alloc() : !hir.memref<64*i32, r>,
                           !hir.memref<64*i32, w>
  %8 = hir.constant 8
  %64 = hir.constant 64

  // Execution of stencilB is overlapped with stencilA.
  hir.call @stencil_A(%Ai, %Bw) at %t
        : (!hir.memref<64*i32, r>, !hir.memref<64*i32, w>)
  hir.call @stencil_B(%Br, %Co) at %t offset %8
        : (!hir.memref<64*i32, r>, !hir.memref<64*i32, w>)
  hir.return
}
\end{lstlisting}

In addition to exploiting parallelism in loops, multiple tasks can be executed
in parallel to further improve performance.
The "task\_parallel" function in listing~\ref{code:hir_task_parallel} shows an
example of task level parallelism expressed in the HIR dialect. Since, the
stencils read the input array and write to the output sequentially, the second
stencil does not have to wait for the first stencil to complete. It can start
its operation as soon as there is enough data to calculate its first output.
After that both the stencil run in lock-steps i.e. in each cycle stencilA
produces one value and stencilB consumes one value. Overlapping the execution of
the tasks reduces the overall latency of the top level function. Like loop
pipelining, overlapped execution of multiple tasks does not add any hardware
overhead either.

\subsection{Loop unrolling}
\begin{lstlisting}[language=MLIR,float,frame=bt,breaklines=true,captionpos=b,
caption={`Unroll loop' in HIR.},label={code:hir_unroll}]
hir.unroll_for %j = 0 to 16 step 1 iter_time(%tj = %t){
  // Start all iterations at the same time instant %t.
  hir.yield at %tj
  // Only unrolled for loops can start all iterations simultaneously.

  // Body.
  ...
}
\end{lstlisting}

Listing~\ref{code:hir_unroll} shows an unroll \textit{for} loop that starts all
its iterations in parallel (since the yield operation happens without any
delay). Unrolling replicates the loop body in hardware. This allows an HIR
design to scale with available hardware resources if there is enough loop
parallelism to exploit. Unrolling can often be combined with pipelining to
further improve parallelism.

HIR does not support partial unrolling of loops. All \textit{unroll\_for} loops
are fully unrolled. Partial unrolling can be represented by stripmining the
\textit{for} loop and completely unrolling the resultant inner loop. This is
very similar to loop vectorization in software compilation. A major difference
is that in software, the vectorizable operations are limited by the CPU, whereas
in a hardware design, arbitrary logic could be replicated to form parallel lanes
of execution.

Both, normal \textit{for} loops and unrolled \textit{for} loops can be
arbitrarily nested within each other. For example, the GEMM benchmark discussed
in Section~\ref{sec:evaluation} uses nested loop unrolling to describe a fully
pipelined, two-dimensional systolic array.

\subsection{Operation chaining, pipelining and retiming}
Operation chaining is the process of scheduling multiple dependent operations in
the same clock cycle. It reduces latency by completing a chain of dependent
operations in the same cycle, which would otherwise span multiple cycles.

The overall performance of the hardware design depends on the maximum frequency
at which the design can reliably operate. Pipelining and retiming help in
reducing delay in the critical path to improve frequency of operation.

Pipelining in hardware design is the process of adding extra registers between
operations in order to break a critical timing path into smaller paths with less
delay. Pipelining a design involves two steps - adding registers between
computations to break them into pipeline stages, and changing the schedule of
the affected operations.  Retiming is another timing closure technique where,
instead of adding new pipeline stages, operations are moved from one pipeline
stage to another. The motivation is to reduce the delay in the critical path at
the expense of non-critical paths. Since, the maximum achievable frequency
depends on the delay of the critical path, this optimization improves the
frequency.

An HIR design uses \texttt{hir.delay} operation to add pipeline registers. All
the above optimizations require adding (pipelining), removing(operator chaining)
or moving the delay operation around. The schedule verifier offers an additional
layer of checks for the validity of these transformations.

\subsection{Memory banking and multi-port RAMs}
\begin{figure}
  \includegraphics[width=0.8\linewidth]{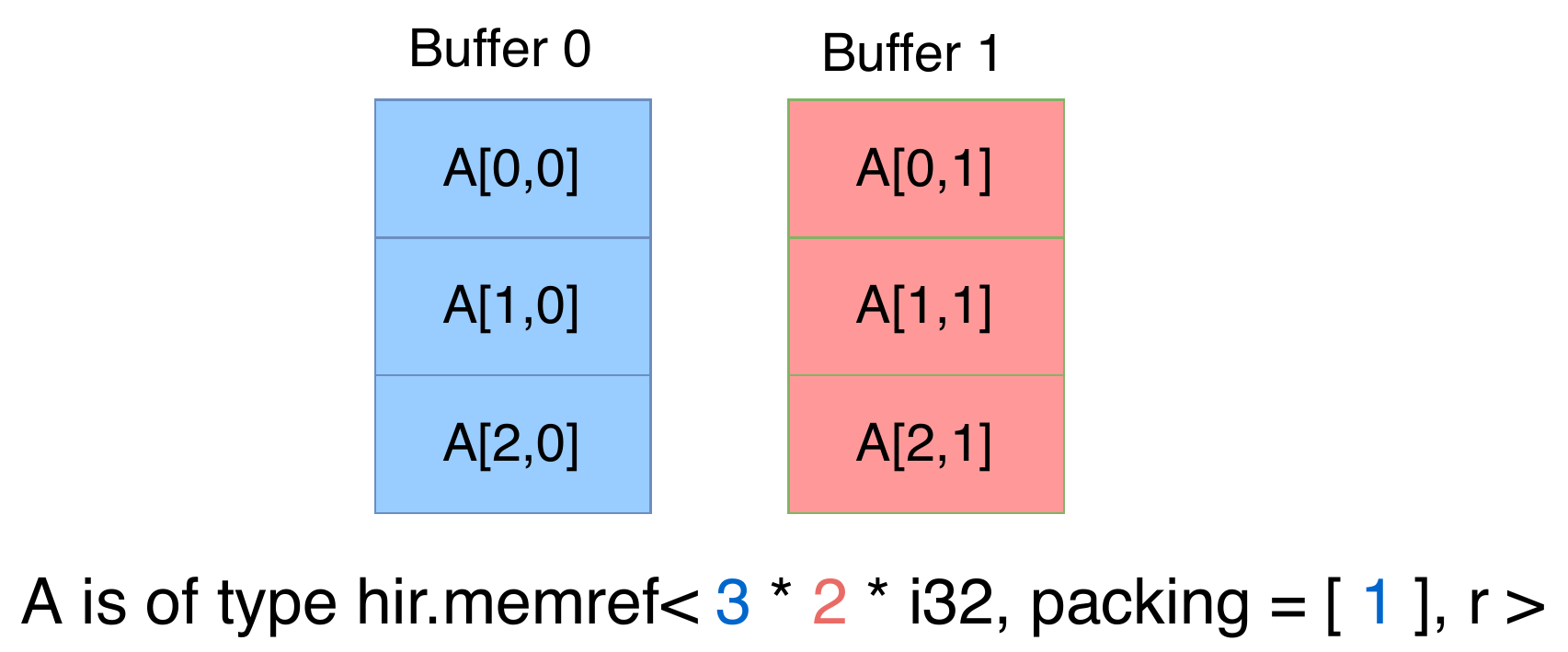}
    \caption{Memory banking in a memref type.}
    \label{fig:memref}
    \vskip -10pt
\end{figure}

Hardware accelerators use on-chip buffers to reduce DRAM accesses. In order to
execute operations in parallel, these designs may need multiple reads and writes
to the buffers every cycle. FPGAs offer multi-ported on-chip RAMs to support
parallel access. The number of parallel ports and the type of port depends on
the the FPGA. An HIR design can instantiate a multi-ported RAM in order to
parallelize memory accesses. Usually the number of ports is limited since adding
multiple ports to a memory requires extra resources in the FPGA.
For workloads where the parallel memory accesses are usually guaranteed to be
separated by a fixed stride, an HIR design may employ memory banking instead.
In this approach, the data is distributed evenly among multiple buffers in such
a way that parallel accesses occur on distinct buffers. HIR supports memory
banking in the memref type. The distributed dimensions of a memref are spread
across multiple buffers whereas addresses that only differ in packed dimensions
fall into the same buffer. Figure~\ref{fig:memref} shows how elements of a
memref with distributed dimensions are spread across multiple buffers.

\section{Evaluation}
\label{sec:evaluation}

\begin{table*}
  \centering
  \begin{tabularx}{0.72\linewidth}{  l c c c c c c c c}
    \toprule
    {Benchmark} & \multicolumn{4}{c}{Vivado HLS/Verilog} &
    \multicolumn{4}{c}{HIR}\\
    \cmidrule(lr){2-5} \cmidrule(lr){6-9} \\
    { }               & LUT & FF & DSP & BRAM & LUT & FF & DSP & BRAM\\
    \midrule
    Matrix transpose & 7    & 51    & 0   & 0 & 8     & 18  & 0     & 0\\
    \midrule
    Stencil-1d       & 152 & 237 & 6 & 0 & 114 & 147 & 6 & 0 \\
    \midrule
    Histogram        & 130 & 107 & 0 & 1 & 101 & 146 & 0 & 1 \\
    \midrule
    GEMM             & 14495 & 24538 & 768 & 0 & 12645 & 29062 & 768 & 0\\
    \midrule
    Convolution      & 1517 & 2490 & 0 & 0 & 289 & 661 & 0 & 0\\
    \midrule
    FIFO (Verilog)    & 34 & 36 & 0 & 1 & 43 & 140 & 0 & 1\\
    \bottomrule
  \end{tabularx}
  \vskip 5pt
  \caption{FPGA resource usage and comparison with Vivado HLS/Verilog.}
  \label{tbl:benchmarks}
  \vskip -10pt
\end{table*}

\begin{table}
  \centering
  \begin{tabularx}{0.95\linewidth}{  l c c r}
    \toprule
    Benchmark        & \multicolumn{2}{c}{Compile times (sec)} & Speedup\\
    \cmidrule(lr){2-3}
    {}               & HIR   & Vivado HLS & {}\\
    \midrule
    Matrix transpose & 0.006 & 13         & 2166$\times$ \\
    \midrule
    Stencil\_1d      & 0.007 & 8          & 1142$\times$ \\
    \midrule
    Histogram        & 0.007 & 13         & 1857$\times$ \\
    \midrule
    GEMM             & 0.099 & 33         & 333$\times$ \\
    \midrule
    Convolution      & 0.013 & 14         & 1076$\times$ \\
    \bottomrule
  \end{tabularx}
  \vskip 5pt
  \caption{Compile times (in seconds) and comparison with Vivado HLS.}
  \label{tbl:compile_times}
  \vskip -10pt
\end{table}

We implemented HIR as an MLIR dialect using mainline MLIR infrastructure.  We
then implemented a translator from the MLIR HIR dialect to Verilog. Our code
is available~\cite{hir-repo}. In order to evaluate the quality of hardware
generated by the HIR compiler, we compared its resource utilization against
Vivado HLS~\cite{vivado}, a proprietary high-level synthesis compiler. Vivado
HLS is the standard for implementing accelerator designs using high-level
synthesis on Xilinx FPGAs.  All results were synthesized for the Xilinx VC709
FPGA evaluation platform. The generated Verilog from both HIR and Vivado HLS
were given to Vivado synthesis tool to obtain the resource usage data. All
results reported are obtained using Vivado and Vivado HLS version 2019.1.

The compile time results reported in Table~\ref{tbl:compile_times} were measured
on a system with 2 x 8-core Intel Xeon Silver 4110 CPUs and 256~GB of DDR4 RAM.
Both the Vivado HLS and HIR designs were synthesized for 200MHz and used the
same amount of loop pipelining and unrolling to match performance.

We compare the quality of generated hardware on four benchmarks.  The matrix
transpose benchmark implements a hardware circuit to read a matrix from an input
memory interface and write the transpose on an output memory interface. The
stencil benchmark takes a one-dimensional array and a set of weights as input
and calculates an output array where each output value is the weighted average
of the inputs within a window. Signal processing operations such as FIR filter
can be classified as a stencil computation. The histogram benchmark calculates a
histogram of an image using a local buffer and writes back the final histogram
to an output memory interface. This benchmark shows data dependent memory
accesses. Matrix multiplication is a fundamental operation in many machine
learning algorithms. We implemented a generalized matrix-matrix multiplication
kernel that reads two 16x16 matrices into buffers, multiplies them using a
systolic array design, and writes back the output. We use distributed RAM to
implement the local buffers for the matrices. All benchmarks operate on 32 bit
integer data.

Table~\ref{tbl:benchmarks} shows the resource utilization of all the designs
implemented in Vivado HLS and in HIR. The HIR compiler generates designs which
have a comparable resource utilization to Vivado HLS, which is a mature high
level synthesis compiler. Our resource usage of DSP blocks is always the same as
Vivado HLS. The results differ only in the utilization of lookup tables and the
number of registers. For matrix transpose, our register usage is much lower than
that of the HLS compiler. We believe this is because the HLS compiler more
aggressively pipelines the design than what is necessary to achieve the desired
200~MHz frequency target. On the GEMM benchmark, the HIR design consumes less
lookup tables but more registers than the HLS design. In stencil-1d HIR performs
better than Vivado HLS in both LUT and register usage by a substantial margin.
In convolution the HIR implementation consumes fewer resources compared to the
HLS implementation for an equivalent design. The FIFO baseline is implemented in
Verilog. The HIR implementation consumes comparable lookup tables but
substantially more registers compared to a Verilog implementation of the FIFO.
Overall, the results in Table~\ref{tbl:benchmarks} show that the HIR compiler
can generate Verilog designs that are comparable to Vivado HLS in resource usage
when both designs have been equally optimized and for the same target frequency.

\section{Related Work}
\label{sec:related-work}

Static Single Assignment (SSA)~\cite{cytron91toplas} based intermediate
representations are a standard in software compilation pipelines. Both
LLVM~\cite{lattner2004llvm} and GCC~\cite{novillo2003treessa} use SSA based IRs
for their compilation pipeline. Some HLS compilers~\cite{canis2011legup,
  pilato2013bambu} also reuse these software IRs for high level synthesis.

\subsection{Hardware intermediate representations}
Like HIR, LLHD~\cite{schuiki2020llhd} is another intermediate representation for
hardware description that was later migrated to MLIR and is hosted as part of
CIRCT~\cite{circt}, an LLVM sub-project and a umbrella initiative to adapt and
use MLIR for high-level synthesis.  LLHD attempts to cover all stages of
hardware synthesis.  It provides IR constructs to describe behavioral and
structural circuits as well as netlists. LLHD is a low-level IR with similar
abstraction level as HDLs (Verilog/VHDL). It does not have {\it for} loops and
it expects that loop unrolling to be done by the language frontend. While LLHD
is suitable for targeting HDLs, we believe that HIR, due to its high level
abstraction, is a better target for HLS and DSL frontends. Currently, HIR lowers
to Verilog, but LLHD could serve as an alternative code generation target. We
intend to contribute HIR to the CIRCT project, and develop it further within its
community.

The FIRRTL~\cite{izraelevitz2017firrtl} IR is designed along with the
Chisel~\cite{bachrach2012chisel} hardware construction language. FIRRTL designs
are represented as an abstract syntax tree. FIRRTL offers features like type and
width inference for easier Chisel interoperability.  Similar to LLHD and in
contrast to HIR, FIRRTL is also a low level IR which can be a suitable target
for HDLs.

$\mu$IR~\cite{sharifian2019uir} decouples the micro-architectural representation
of the accelerator from its behavioral specification.
Mircroarchitectural optimizations like pipelining and retiming are implemented
as transformations of structural graph.
Futil~\cite{anonymized2020futil} represents an accelerator using a structural
sub-language and a control sub-language. The control sub-language describes a
coarse grained schedule via a {\it happens-before} relation.
On the other hand, HIR captures the complete accelerator description through a
unified representation.  One of the use cases of HIR is to allow expert
developers to hand optimize kernels. We believe that a unified representation is
easier to understand, optimize and debug.  Unlike Futil, HIR's schedules are
exact (cycle accurate). This helps in interfacing with external Verilog modules
without the overhead of handshaking.  The HIR compiler also helps in detecting
errors like pipeline imbalance after microarchitectural optimizations.

\subsection{HIR for HLS compilation}
In this section, we briefly discuss high-level synthesis languages and how
HIR could fit in as an IR in their compilation flow.
Bluespec-Verilog~\cite{nikhil2004bluespec} represents a circuit as a set of
atomic rules. A recent work~\cite{bourgeat2020koika} attempts to enhance the
language for user-level control over the schedule and predictable performance.
We believe that HIR's ability to express a design at a higher level of
abstraction than RTL or structural level while providing complete control over
the schedule to allow fine grained task level parallelism would be a good fit as
a target intermediate language for the BSV compiler.
The Dahlia\cite{nigam2020dahlia} language is inspired from the observation that
HLS languages generate unpredictable designs due to their excessive flexibility.
A small change in the design can affect the performance by a large factor.
Dahlia and its affine type system enforces the high level design to respect the
limitations of hardware. For example, a valid design is guaranteed to never have
multiple reads and writes on the same memory in the same cycle. Many of Dahlia's
language features such as memory banking, for loops and unroll-for loops have a
direct equivalent in HIR. This would make it very easy to use HIR as the
intermediate representation in its compiler. Also, a valid Dahlia design does
not introduce undefined behaviour in HIR. For an IR to use static analysis to
prove the absence of undefined behaviour would be too conservative since it may
be targeted by a variety of frontend languages. Thus, the ideal situation is
when the language is conservative in accepting a design and the IR is flexible
to cater to the need of multiple languages.
Aetherling~\cite{durst2020aetherling} is another interesting work on high level
synthesis. It introduces space-time types that can be used to describe dataflow
pipelines with exact throughput and latency requirements. The language attempts
to remove the need for synchronization overhead between different stages of the
pipeline. HIR is an ideal fit for such a language. The explicit deterministic
scheduling completely removes the need for extra synchronization stages. HIR can
represent designs where the producer and consumer operate in lock steps and there
is no back-pressure between the stages and Aetherling's type system ensures that
this would always be the case.

We believe that the combination of high level language constructs such as
{\it for} and {\it unroll-for} loops, the ability to specify exact schedules of
operations and the ability to represent both high level optimizations (such as
loop pipelining and task overlapping) as well as low level optimizations such as
retiming, allows HLS compilers to use HIR during the whole compilation pipeline.

\section{Conclusion}
We introduced HIR, an intermediate representation to describe FPGA based
hardware accelerator designs. HIR has been implemented as a dialect in the MLIR
compiler infrastructure.  It captures the hardware design at a higher level of
abstraction than today's hardware description languages like Verilog or VHDL.
This makes it easier for DSL frontends to target HIR for code generation.
HIR is built in a rigorous manner drawing from best practices learnt from
compiler IR design in the open-source communities of LLVM and MLIR. The HIR
language simplifies the process of hardware optimization and unifies the
hardware compilation pipeline with general-purpose software compiler
infrastructure to share core infrastructure, utilities and passes maximally.
Preliminary results demonstrate the effectiveness of the approach in terms
of hardware designers' productivity, efficiency of the code generated, and code
generation time improvements of over 1000$\times$ over Xilinx Vivado HLS.

\bibliography{references}



\end{document}